\documentclass[11pt]{article}
\usepackage[dvips]{graphicx}
\def\be{\begin{equation}}
\def\ee{\end{equation}}
\def\bea{\begin{eqnarray}}
\def\eea{\end{eqnarray}}
\def\l#1{\overline{#1}}
\def\b#1{\beta_{#1}}
\def\g#1{\gamma_{#1}}
\def\du{\mu^2\frac{\partial}{\partial\mu^2}}
\def\f#1#2{\frac{#1}{#2}}

\newcommand{\qslash}{q \!\!\! /}

\begin{document}
{\begin{center} {\LARGE {Avoiding the Landau-pole in perturbative
QCD} } \\
[8mm] {\large K. Van
Acoleyen\footnote{karel.vanacoleyen@rug.ac.be} and H. Verschelde}

\end{center}
\noindent Department of Mathematical Physics and Astronomy,
University of Ghent, Krijgslaan 281 (S9), 9000 Ghent, Belgium.

\vspace{5cm} \noindent {\bf Abstract.}We propose an alternative
perturbative expansion for QCD. All scheme and scale dependence is
reduced to one free parameter. Fixing this parameter with a
fastest apparent convergence criterion gives sensible results in
the whole energy region. We apply the expansion to the calculation
of the zero flavor triple gluon vertex, the quark gluon vertex,
the gluon propagator and the ghost propagator. A qualitative
agreement with the corresponding lattice results is found.
\newpage
\section{Introduction}
Perturbation theory is by far the most successful tool to get
quantitative predictions from a field theory. Unfortunately, the
results depend on the renormalization scale and scheme and the
number of free parameters describing this dependence grows with
the order of truncation. In most cases one does not bother too
much about this dependence and simply chooses a scheme ($\l{MS}$,
MOM,$\ldots$) which is supposed to give good results. More
sophisticated approaches select a different \textit{ideal} scheme
for each perturbative series. Most cited in this context are
Stevenson's principle of minimal sensitivity \cite{s81} and
Grunberg's  method of effective charges \cite{g84}.

In this paper we will reorganize the conventional perturbation
series in a new alternative expansion, which has only one
redundant free parameter, the expansion parameter itself. This
parameter will be fixed by a fastest apparent convergence
criterion (facc). We find similar results as for the ordinary QCD
perturbation theory in the UV-region and for intermediate
energies. In the IR-region on the contrary, our expansion still
gives sensible results whereas the \textit{normal} perturbation
theory becomes useless if one approaches the Landau-pole.

In section \ref{rewriting} we will first review some of the major
aspects of the ordinary perturbation theory and then use this as
starting point for the alternative expansion. Some results will be
presented in section \ref{results}. We perform calculations on the
triple gluon vertex, the quark gluon vertex, the gluon propagator
and the ghost propagator. Our results are compared with the
conventional perturbation theory and with the corresponding
lattice results. We also show that the reason for the IR finite
results lies in a peculiar behavior of the running expansion
parameter $y$, resulting from the facc. This behavior should be
contrasted with the running coupling $\alpha$ that Shirkov and
Solovtsov obtain by imposing analyticity \cite{ss97}. While we
find a universal power behavior for $y$, they find an IR-finite
value for $\alpha$ at zero momentum.
\section{Rewriting perturbation theory} \label{rewriting}
In the following we will consider the perturbative calculation of
a renormalization scheme and scale invariant quantity
$\mathcal{R}$, that is function of only one external scale $q^2$,
in a massless version of QCD . In ordinary perturbation theory one
finds a row of approximations $\mathcal{R}^n$ for $\mathcal{R}$,
with \be
\mathcal{R}^n={h^{(n)}}^N(1+r_{1}(q^2)h^{(n)}+r_2(q^2){h^{(n)}}^2
+\ldots+r_n(q^2){h^{(n)}}^n)~. \label{R}\ee With $N$ depending on
the calculated quantity. The coupling constant $h^{(n)}$ is the
solution of \bea
\b{0}\ln\f{\mu^2}{\Lambda^2}&=&\f{1}{h}+\f{\b{1}}{\b{0}}\ln(\b{0}h)
+\nonumber\\&&\int^h_0\!\!\!dx\left(\f{1}{x^2}-\f{\b{1}}{\b{0}}\f{1}{x}
-\f{\b{0}}{\b{0}x^2+\b{1}x^3+\ldots+\b{n}x^{n+2}}\right),~~~
\label{coupling} \eea with $ \du h \equiv
-(\b{0}h^2+\b{1}h^3+\b{2}h^4+\ldots).$ As mentioned in the
introduction, every truncation $\mathcal{R}^n$ is highly scale and
scheme dependent. One can for instance describe this dependence
with the free parameters \cite{s81} $\b{0}\ln\f{\mu^2}{\Lambda^2},
\b{2}, \b{3}, \ldots, \b{n}.$ The dependence of the coefficients
$r_i$ on each of these parameters, cancels the dependence of the
coupling constant $h^{(n)}$ up to order ${h^{(n)}}^{n+1}$.

In many cases one simply chooses a scheme ($\l{MS}$, MOM,
$\ldots$) and sets the scale $\mu^2$ equal to the external scale
$q^2$. To get reliable results, one must hope that the first,
second, or third order approximation lies \textit{close enough} to
the exact result. The working hypothesis of perturbation theory is
that the perturbation series is asymptotic to the exact result
$\mathcal{R}$ \cite{b98}. One then obtains an error estimation, by
assuming that \bea
\left|\f{\mathcal{R}-\mathcal{R}^n}{\mathcal{R}}\right|
\sim\left|\f{\mathcal{R}^{n+1}-\mathcal{R}^n}{\mathcal{R}^{n+1}}\right|
\equiv\Delta^n~~. \label{delta} \eea

Other approaches fix the scale/scheme by imposing a condition on
the truncated series. The minimal sensitivity condition \cite{s81}
gives a different scale/scheme for every approximation. The method
of the effective charges \cite{g84}, sets $q^2=\mu^2$, the free
parameters are then obtained by demanding the coefficients $r_{i}$
to be zero. One now also finds $\Delta^n$ to give a good
estimation of the error \cite{ms94}.

As a consequence of asymptotic freedom, conventional perturbation
theory works well in the UV-region, which is reflected in low
values of $\Delta^n$ (n=1,2,3) for high values of $q^2$.
Unfortunately $\Delta^n$ gets larger if we lower the external
scale $q^2$. This signals that one has to add non-perturbative
power corrections to the conventional perturbation theory
\cite{b98}. For intermediate energies, the sum rules \cite{svz79}
successfully relate many of these power corrections to a few
condensates. A further lowering of $q^2$ towards the IR-region is
catastrophic in most cases \footnote{Exceptions can be found in
\cite{ms94}, where the minimal sensitivity criterion selects a
scheme with an IR fixed point.}, $\Delta^n$ diverges together with
${\mathcal{R}}^n$ as one encounters the Landau-pole and
perturbation theory becomes useless.

The alternative expansion we propose in this paper has no
Landau-pole problem and gives sensible results up to $q^2=0$ with
a reasonable error estimation. We simply have to exchange $h$ for
a new expansion parameter $y$, defined
by\be\b{0}\ln\f{\mu^2}{\Lambda^2}\equiv\f{1}{y}+\f{\b{1}}{\b{0}}\ln(\b{0}y).\label{defy}
\ee From (\ref{R}),(\ref{coupling}) and (\ref{defy}) we find after
some calculations \bea \mathcal{R}&=&y^N
\Bigg(1+y\bigg[A_{1}+Nk\bigg]
+y^2\bigg[A_{2}+N\Big(\f{\b{2}}{\b{0}}-(\f{\b{1}}{\b{0}})^2\Big)\nonumber\\
&&+k\Big((N+1)A_{1}+N\f{\b{1}}{\b{0}}\Big)+k^2\f{N}{2}(N+1)\bigg]\nonumber\\
&&+y^3\bigg[A_{3}+A_{1}(N+1)\Big(\f{\b{2}}{\b{0}}-(\f{\b{1}}{\b{0}})^2\Big)
+\f{N}{2}\Big(\f{\b{3}}{\b{0}}-(\f{\b{1}}{\b{0}})^3\Big)\nonumber\\
&&+k\Big(A_{1}(N+1)\f{\b{1}}{\b{0}}+(N+2)A_{2}+N(N+2)\f{\b{2}}{\b{0}}\label{expansion}\\
&&-N(N+1)(\f{\b{1}}{\b{0}})^{2}\Big)\nonumber\\
&&+k^2\Big(\f{A_{1}}{2}(N+2)(N+1)+N(N+\f{3}{2})\f{\b{1}}{\b{0}}\Big)\nonumber\\
&&+k^3\Big(\f{N}{6}(N+2)(N+1)\Big)\bigg]+\ldots\Bigg)\nonumber\eea
with \bea
k&\equiv&k(q^2,y)=\f{1}{y}+\f{\b{1}}{\b{0}}\ln(\b{0}y)-\b{0}\ln\f{q^2}{{\Lambda_{\l{MS}}}^2}\\
A_{i}&\equiv&r_{i}(\mu^2=q^2,\l{MS})  \eea The
$\beta$-coefficients $\b{2},\b{3},\ldots$ are also in the
$\l{MS}$-scheme.

We now find a row of approximations $\mathcal{R}^{n}(y)$ with all
the redundant dependence residing in one single free parameter $
\b{0}\ln\f{\mu^2}{\Lambda^2}$ or equivalently $y$. All the other
scheme dependence has disappeared, because it was eliminated from
the expansion parameter. This might seem strange, since we
explicitly refer to the $\l{MS}$ scheme, but one can show that
another choice for the reference scheme changes the coefficients
$A_{i}$\footnote{ The scheme/scale dependence of
$r_{1},r_{2},r_{3}$ was derived in \cite{kss84}.} and $\b{i}$ in
such a way that it exactly compensates the shift of
$k~(k\rightarrow k'=k+2\b{0}\ln\f{\Lambda'}{\Lambda_{\l{MS}}})$.

To fix the expansion parameter $y$ for each value of $q^2$, we
will have to impose some condition. We choose a facc: for each
approximation $\mathcal{R}^{n}(y)$ we will set $y$ equal to $y_n$,
the expansion parameter that minimizes the relative correction to
the first order truncation of the series: \bea \min
\left|\f{\mathcal{R}^{n}(y)-\mathcal{R}^{1}(y)}{\mathcal{R}^{1}(y)}\right|
=\left|\f{\mathcal{R}^{n}(y_{n})
-\mathcal{R}^{1}(y_{n})}{\mathcal{R}^{1}(y_{n})}\right|
\label{facc}\eea We will use the same error estimation as for the
conventional perturbation theory: \bea
\Delta^n\equiv\left|\f{\mathcal{R}^{n+1}(y_{n+1})-
\mathcal{R}^{n}(y_{n})}{\mathcal{R}^{n+1}(y_{n+1})}\right|
\label{deltafacc}\eea (Notice the difference: the fixing of y is
done on the series, while the error is estimated on the results
obtained after fixing.) As for any other possible fixing condition
(e.g. minimal sensitivity, another facc,...), there is no rigorous
mathematical motivation for the condition we use. The true
motivation lies in the fact that it generates sensible results
with a good error estimation. This is the case for all the
calculations we have performed so far. If we use, for example, the
more obvious facc, where one minimizes the relative correction to
the zero order truncation $y^N$, we do not find sensible results
for the whole range of energies. There is a discontinuity at the
point that separates the high energy region where this correction
minimizes to zero, and the low energy region where it minimizes to
a nonzero value. Let us now present some results obtained from
(\ref{expansion}) and (\ref{facc}).
\section{Some results}\label{results}
We now demonstrate the $y$-expansion on some quantities that also
have been calculated on the lattice. All the needed two and three
loop results have been calculated by Chetyrkin and Retey
\cite{cr00}. Everything is in Landau gauge for $N_{c}=3$ and
$N_{f}=0$. We will take the method of effective charges to be
exemplary for the ordinary perturbation theory, but similar
results are found with any other approach to the conventional
perturbation theory.
\subsection{Triple gluon vertex}
There are several ways in which one can associate a
(renormalization) scale and scheme invariant coupling constant
with the triple gluon 3-point function \be
G^{(3)abc}_{\mu\upsilon\rho}(p,q)\equiv
i^2\int\!\!\!dxdye^{-i(px+qy)}<T[A_{\mu}^{a}(x)A_{\upsilon}^{b}(y)A_{\rho}^{c}(0)]>,
 \ee or more precisely with its related vertex function
 $\Gamma_{\mu\upsilon\rho}^{abc}(p,q,-p-q)$, defined by \be
 G^{(3)abc}_{\mu\upsilon\rho}(p,q)\equiv
 D_{\mu\mu'}^{ad}(-p)D_{\upsilon\upsilon'}^{be}(-q)D_{\rho\rho'}^{cf}(-p-q)
 \Gamma_{\mu'\upsilon'\rho'}^{def}(p,q,-p-q),
\ee where \be D_{\mu\upsilon}^{ab}(q)\equiv
i\int\!\!\!dxe^{iqx}<T[A_{\mu}^{a}(x)A_{\upsilon}^{b}(0)]>. \ee If
one sets one external momentum to zero, one finds \cite{cr00} that
the vertex-function can be written as: \bea
\Gamma_{\mu\upsilon\rho}^{abc}(q,-q,0)=-igf^{abc}\Big((2g_{\mu\upsilon}q_{\rho}-
g_{\mu\rho}q_{\upsilon}-g_{\rho\upsilon}q_{\mu})T_1(q^2)\Big)\nonumber\\
-(g_{\mu\upsilon}-\f{q_{\mu}q_\upsilon}{q^2})q_{\rho}T_{2}(q^2)\Big)\eea
The coupling that was calculated on the lattice \cite{bbr00,blm98}
is found to be (\cite{cr00},section 6.4):\be
\alpha_{s}(q^2)\equiv4\pi
h^{\widetilde{{\textrm{{MOM}}}}\textrm{gg}}(q^2)=h\Big(T_{1}(-q^2)-\f{1}{2}T_{2}(-q^2)\Big)^2Z(-q^2)^3,\ee
 \bea \textrm{where}\qquad\qquad h&=&\f{g^2}{16\pi^2},\\
D_{\mu\upsilon}^{ab}(q)&=&\delta^{ab}(g_{\mu\upsilon}-\f{q_{\mu}q_\upsilon}{q^2})\f{Z(q^2)}{q^2}.\label{gluonprop}
\eea One can easily check the scheme and scale independence of
$\alpha_{s}$. The three-loop result for
$h^{\widetilde{\textrm{MOM}}\textrm{gg}}$ in the $\l{MS}$-scheme
for $\mu^2=q^2$ is,\cite{cr00}:\bea
h^{\widetilde{\textrm{MOM}}\textrm{gg}}&=&h+h^2[\f{70}{3}]+h^3[\f{516217}{576}-\f{153}{4}\zeta_{3}]+\nonumber\\
&&h^4[\f{304676635}{6912}-\f{299961}{64}\zeta_{3}-\f{81825}{64}\zeta_{5}],\label{hmomgg}\eea
where $\zeta_{i}$ is the Riemann zeta function. From this we can
read of the coefficients $A_{1},A_{2},A_{3}$ that are needed in
(\ref{expansion}). The $\beta$-coefficients in the $\l{MS}$-scheme
have been calculated up to four loops in \cite{vvl97}:\bea
\b{0}=11,\qquad\b{1}=102,\qquad\b{2}=\f{2857}{2},\qquad\b{3}=\f{149753}{6}+3564\zeta_{3}.\label{betams}\eea
We will compare our two and three loop results for
$\alpha_{s}(q^2)$ obtained from (\ref{expansion}) (with $N=1$) and
(\ref{facc}) with the results obtained from the method of
effective charges \cite{g84} or equivalently in the
$\widetilde{\textrm{MOM}}$gg-scheme defined on the triple gluon
vertex. The two and three loop  $\widetilde{\textrm{MOM}}$-scheme
results are found as the solution of (\ref{coupling}) with
$n=2,3$. The $\widetilde{\textrm{MOM}}$-scheme
$\beta$-coefficients can be easily obtained from (\ref{hmomgg})
and (\ref{betams}):\bea
\b{2}^{\widetilde{\textrm{MOM}\textrm{gg}}}&=&\f{186747}{64}-\f{1683}{4}\zeta_{3},\nonumber\\
\b{3}^{\widetilde{\textrm{MOM}\textrm{gg}}}&=&\f{20783939}{128}-\f{1300563}{32}\zeta_{3}-\f{900075}{32}\zeta_{5}.\eea
While the $\Lambda$ parameter is given by \cite{cg79}:\be
2\b{0}\ln\f{\Lambda_{\widetilde{\textrm{MOM}\textrm{gg}}}}{\Lambda_{\l{MS}}}=\f{70}{3}.\ee
Our two and three loop results are plotted together with the two
and three loop $\widetilde{\textrm{MOM}}$-results in figs.
\ref{3gl} and \ref{3gli}.
\begin{figure}
\begin{center}
\includegraphics[angle=-90,width=10cm]{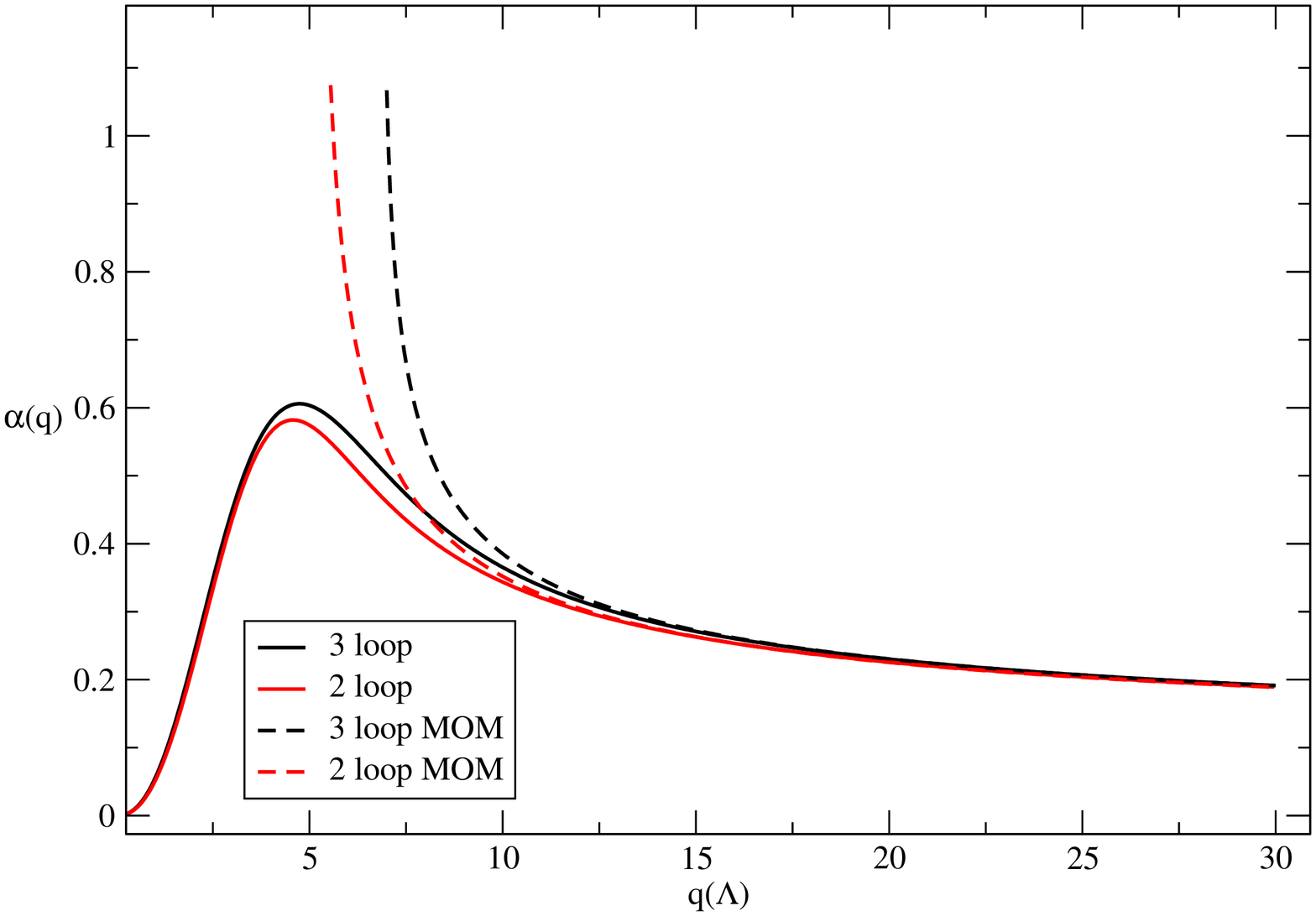}
\caption{$\alpha_{s}(q)$ ($q$ in units of
$\Lambda_{\l{\textrm{MS}}})$, for 2 and 3 loops in the
$y$-expansion and in the $\widetilde{\textrm{MOM}}$-scheme.}
\label{3gl} \vspace{1cm}
\includegraphics[angle=-90,width=11cm]{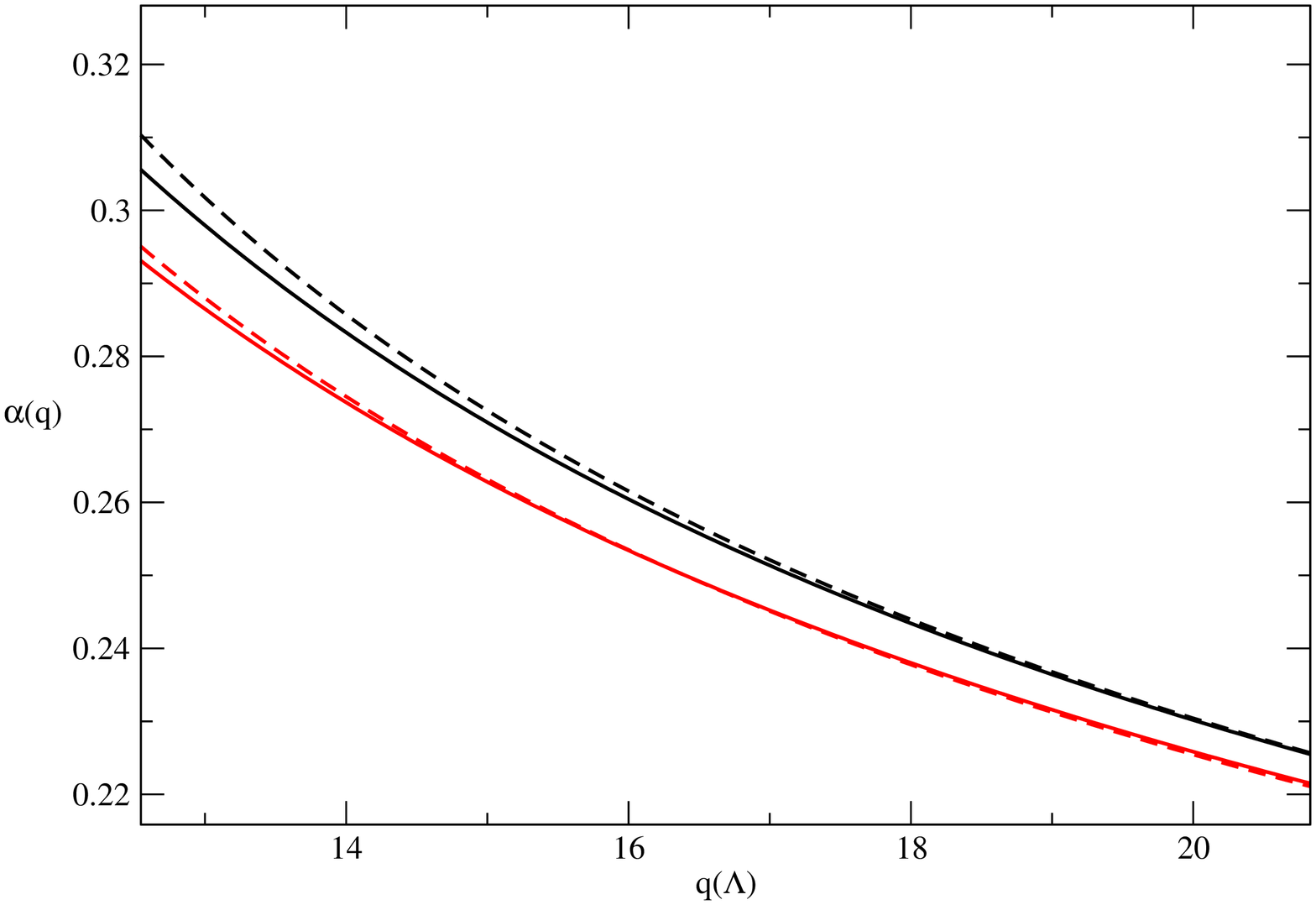}
\caption{Zooming in on the intermediate energy region of fig.
\ref{3gl}} \label{3gli}
\end{center}
\end{figure}

We can clearly distinct three regions. For
$q>30\Lambda_{\l{\textrm{MS}}}$ one finds the UV-region: the four
results for $\alpha_{s}$ coincide and the perturbation theory is
completely reliable. The intermediate energies region goes from
$q\approx 30\Lambda_{\l{\textrm{MS}}}$ down to $q\approx
10\Lambda_{\l{\textrm{MS}}}$. A difference grows between the two
and three loop results, but for both orders the $y$-expansion
results still coincide with the $\widetilde{\textrm{MOM}}$
results. Power corrections are expected. For
$q<10\Lambda_{\l{\textrm{MS}}}$ we find ourselves in the
IR-region. The $\widetilde{\textrm{MOM}}$ results diverge while
the $y$-expansion results continue to behave in a sensible way.

\begin{figure}
\begin{center}
\includegraphics[angle=-90,width=10cm]{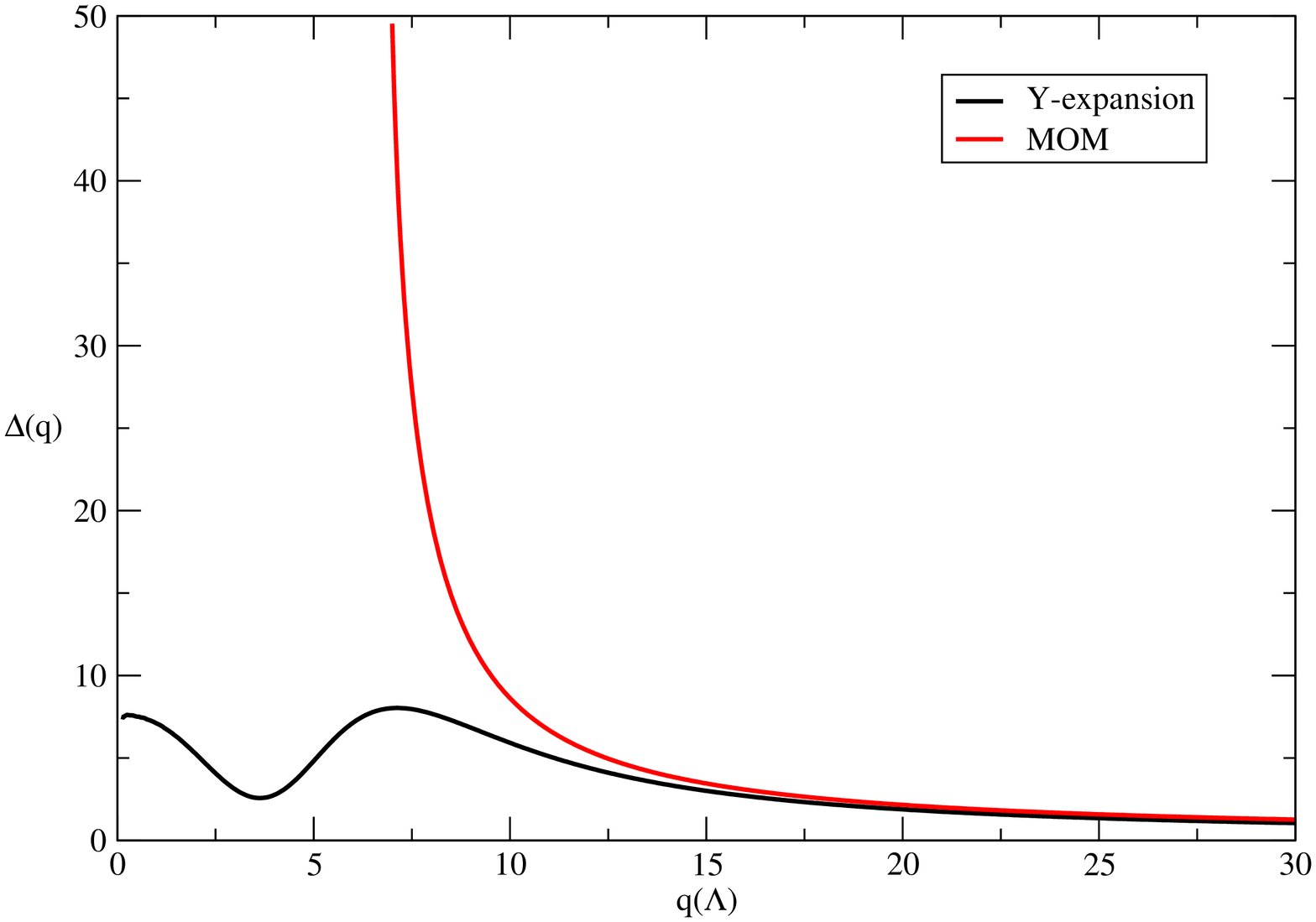}
\caption{$q$ (in units of $\Lambda_{\l{\textrm{MS}}})\rightarrow
\Delta^{2}=\left|\f{\alpha_{s}^{(2)}-\alpha_{s}^{(3)}}{\alpha_{s}^{(3)}}
\right|$ (in $\%$), for both the $y$-expansion and the
$\widetilde{\textrm{MOM}}$ scheme. } \label{deltaf}
\end{center}
\end{figure}

The same conclusions can be read from fig. \ref{deltaf}, where
$\Delta^2$ (see (\ref{delta})) is plotted, both for the
$\widetilde{\textrm{MOM}}$ scheme and for the $y$-expansion. In
the IR-region the error estimation diverges for the
$\widetilde{\textrm{MOM}}$ scheme, while it stays in an acceptable
interval for the $y$-expansion.

It is the facc (\ref{facc}) that keeps the error estimation under
control in the IR. This criterion, and in fact every other
sensible criterion, will select for each momentum $q$ a value for
y that makes the higher order ($n>1$) terms in the series
(\ref{expansion}) as small as possible. Both for small values of
$q$ ($q\ll\Lambda_{\l{\textrm{MS}}}$) and for large values
($q\gg\Lambda_{\l{\textrm{MS}}}$) it is the value of \be
yk(q^2,y)=1+y\f{\b{1}}{\b{0}}\ln(\b{0}y)-y\b{0}\ln\f{q^2}{{\Lambda_{\l{MS}}}^2},
\ee that determines the size of these higher order terms. For
$q\gg\Lambda_{\l{\textrm{MS}}}$ the large logarithm will be
compensated by the $y$ that multiplies it. One finds the usual
high energy running of the expansion parameter:\be
y\stackrel{q\rightarrow\infty}{=}\f{1}{\b{0}\ln\f{q^2}{\Lambda_{\l{MS}}^2}+c}~~~,\label{yhigh}
\ee with $c$ a constant. This gives \be
k(q^2,y)y\stackrel{q\rightarrow\infty}{\approx}\f{1}{\b{0}\ln\f{q^2}{\Lambda_{\l{MS}}^2}}\left(c-\f{\b{1}}{\b{0}}\ln(\ln\f{q^2}{\Lambda_{\l{MS}}^2})\right).\ee
For $q\ll\Lambda_{\l{\textrm{MS}}}$ the same cancellation can not
occur since $y$ must be positive, the large logarithm will now be
compensated by the logarithm in the second term, we find a power
behavior for $y$: \be
y\stackrel{q\rightarrow0}{=}c'\left(\f{q^2}{\Lambda_{\l{MS}}^2}\right)^{\f{\b{0}^2}{\b{1}}}\label{ylow}\ee
and \be
k(q^2,y)y\stackrel{q\rightarrow0}{\approx}1\label{klow}.\ee

This high and low energy behavior of the expansion parameter $y$
is completely universal, it is independent of the order of
truncation, of the coefficients $A_{i}$ and to a certain extent of
the criterion that was used. The running of $y$ is depicted in
fig. \ref{y} together with the fitted low (\ref{ylow}) and high
(\ref{yhigh}) energy behavior for the three loop truncation.
\begin{figure}
\begin{center}
\includegraphics[angle=-90,width=10cm]{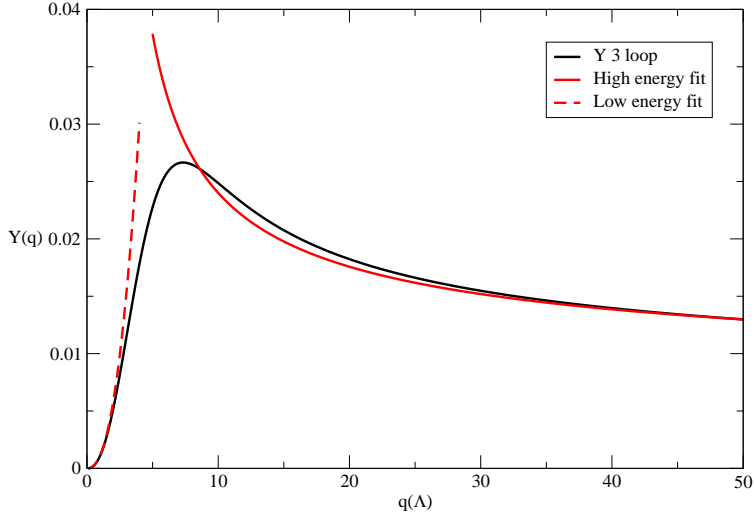}
\caption{$y(q)$ ($q$ in units of $\Lambda_{\l{\textrm{MS}}}$) for
the tree loop truncation. Also depicted: the low and high energy
fits (\ref{ylow}) and (\ref{yhigh}).} \label{y}
\end{center}
\end{figure}

If one would use the series itself to estimate the truncation
error, (\ref{klow}) would seem to invalidate the expansion for low
energies, since the higher order terms become order 1. However, if
you look at the the row of truncations (\ref{deltafacc}) to
estimate the error , the expansion remains valid (at least for low
orders) since $2.5\%<\Delta^2<7.5\%$ (see fig. \ref{delta}). We
have found a same behavior of $\Delta^2$ for every other possible
vertex-coupling that could be calculated from \cite{cr00}.

\begin{figure}
\begin{center}
\includegraphics[angle=-90,width=11cm]{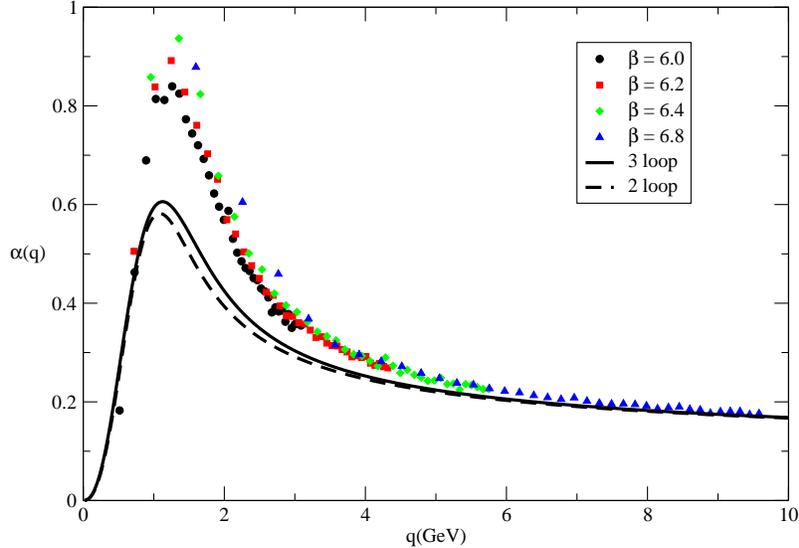}
\caption{The lattice results \cite{bbr00} for the coupling from
the triple gluon vertex with the two and three loop results of the
y-expansion for $\Lambda_{\l{MS}}$=237 Mev.} \label{lat3gl}
\end{center}
\end{figure}

We will finally compare our results with the lattice results of
\cite{bbr00}. This requires a fit of $\Lambda_{\l{MS}}$, which was
done for the two and three loop $\widetilde{\textrm{MOM}}$ results
in \cite{bbr00} and \cite{cr00} in the intermediate energy region
(3 Gev-10 Gev). It was found that the
$\widetilde{\textrm{MOM}}$-results could be best fitted to the
lattice results if a power correction $\f{c}{p^2}$ was added. The
fitted two and three loop values of $\Lambda_{\l{MS}}$ are: 235
Mev and 238 Mev. The three loop power correction is 30$\%$ less
then the two loop one.

Since in the intermediate energy region the results of the
$y$-expansion are the same as the $\widetilde{\textrm{MOM}}$
results we can rely on the aforementioned fits. We will use the
same value $\Lambda_{\l{MS}}$=237 Mev, for every order. The two
and three loop results of the $y$-expansion are plotted together
with the lattice results in fig. \ref{lat3gl}. As expected, the
difference between our results and the lattice result can be
fitted as a power correction for $q>3$ Gev. The amplitude of our
maximum is significantly smaller then the amplitude for the
lattice maximum in the IR region. But both maxima seem to approach
each other, our amplitude grows larger with the order of
truncation while the lattice amplitude becomes smaller for larger
volumes (smaller $\beta$).
\subsection{The quark gluon vertex}
Again, there are several ways one can associate a scale and scheme
invariant running coupling with the (zero flavor\footnote{no
internal fermion loops}) quark-gluon vertex $\Lambda_{\mu
ij}^{a}$, which is defined by:\bea G_{\mu
ij}^{(3)a}(p,q)=S_{ii'}(-p)\Lambda_{\mu'
i'j'}^{d}(p,q,-q-p)S_{j'j}(q)D_{\mu'\mu}^{ad}(p+q)~~~,\eea where
$G_{\mu ij}^{(3)a}$ is the corresponding 3-point function and
$S_{ij}$ is the quark propagator. After setting the external gluon
momentum equal to zero, the vertex can be written as,
\cite{cr00}:\be \Lambda_{\mu
ij}^{a}(-q,q,0)=gT_{ij}^{a}\left[\g{\mu}\Lambda_{g}(q^2)
+\g{\upsilon}\left(g_{\mu
\upsilon}-\f{q_{\mu}q_{\upsilon}}{q^2}\right)\Lambda_{g}^{T}(q^2)
\right]\ee We find the coupling constant that was defined and
calculated on the lattice in \cite{s98,skw01} to be: \be g(q^2)=
4\pi
h^{\f{1}{2}}\left(\Lambda_{g}(-q^2)+\Lambda_{g}^{T}(-q^2)\right)Z^{\f{1}{2}}(-q^2)Z_{2}(-q^2)~~,\ee
where \be S_{ij}(q)=-\delta_{ij}\f{\qslash}{q^2}Z_2(-q^2). \ee
From \cite{cr00} one finds, with $\mu^2=q^2$ in the
$\l{\textrm{MS}}$ scheme: \bea g(q^2)&=&4\pi
h^{\f{1}{2}}\Big(1+h[\f{151}{24}]+h^2[\f{87557}{384}-47\zeta_{3}]+\nonumber\\
&&h^3[\f{266866067}{27648}-\f{824999}{288}\zeta_{3}-\f{349225}{1152}\zeta_{5}]+\ldots\Big)\label{g}\eea
Putting these coefficients, together with the $\beta$-coefficients
(\ref{betams}) in (\ref{expansion}) (now for $N$=1/2) and fixing
$y$ with the facc (\ref{facc}) will give us the two and three loop
$y$-expansion results for $g(q^2)$.

The two and three loop $\widetilde{\textrm{MOM}}$ scheme results
can now be found as $4\pi (h^{(n)})^{\f{1}{2}}$ (n=2,3), with
$h^{(n)}$ solution of (\ref{coupling}). From (\ref{betams}) and
(\ref{g}) we can easily determine the needed $\beta$-coefficients:
\bea
\b{2}^{\widetilde{\textrm{MOM}}\textrm{qg}}&=&\f{185039}{48}-1034\zeta_{3}~~,\nonumber\\
\b{3}^{\widetilde{\textrm{MOM}}\textrm{qg}}&=&\f{32456317}{192}-\f{4134361}{72}\zeta_{3}
-\f{3841475}{288}\zeta_{5}~~.\eea The $\Lambda$-parameter is now
given by:\be
2\b{0}\ln\f{\Lambda_{\widetilde{\textrm{MOM}\textrm{qg}}}}{\Lambda_{\l{MS}}}=\f{151}{12}.\ee
\begin{figure}
\begin{center}
\includegraphics[angle=-90,width=10cm]{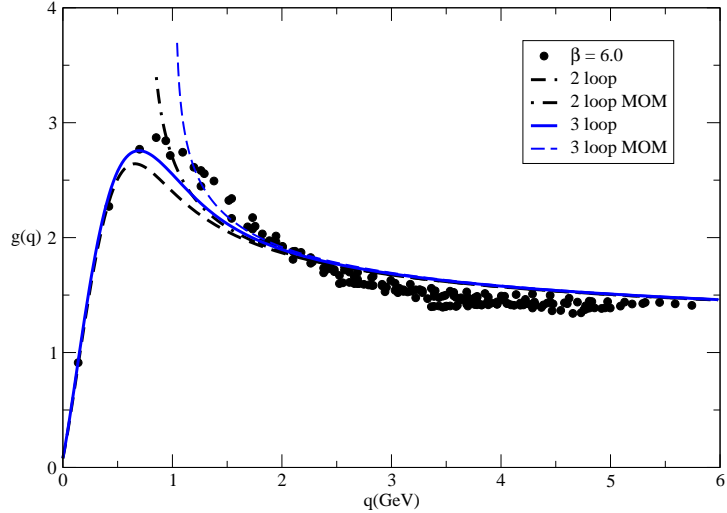}
\caption{The lattice results \cite{skw01} for the coupling from
the quark gluon vertex with the two and three loop results from
the $y$-expansion and the $\widetilde{\textrm{MOM}}$ scheme, with
$\Lambda_{\l{MS}}$=237 Mev. } \label{qugl}
\end{center}
\end{figure}
The results are completely similar to the results for the triple
gluon vertex. Instead of performing a separate fit we simply take
the value ($237$ Mev) for $\Lambda_{\l{MS}}$ obtained from the
triple gluon vertex, to compare with the lattice result. From fig.
\ref{qugl} we can again observe a turnover for the $y$-expansion
results and the lattice results around 1 Gev, where the
$\widetilde{\textrm{MOM}}$ scheme results diverge. We finally note
that the lattice results were obtained for a (small) non-zero
quark mass, while our results are for a massless quark, so we
should not be too enthusiastic about the small amplitude
difference in the IR-region.
\subsection{The gluon propagator}
The $y$-expansion will now be applied to the calculation of the
scale and scheme invariant gluon propagator
$\widehat{D}_{\mu\upsilon}^{ab}(-q^2)$ defined by:\be
\widehat{D}_{\mu\upsilon}^{ab}(-q^2)\equiv
f(h)D_{\mu\upsilon}^{ab}(-q^2),  \ee with \be
\mu^2\f{\partial}{\partial\mu^2}D_{\mu\upsilon}^{ab}(-q^2)
\equiv(\g{3_{0}}h+\g{3_{1}}h^2+\g{3_{2}}h^3+\ldots)D_{\mu\upsilon}^{ab}(-q^2)
\ee and \be
\mu^2\f{\partial}{\partial\mu^2}f(h)\equiv-(\g{3_{0}}h+\g{3_{1}}h^2+\g{3_{2}}h^3+\ldots)f(h).\label{f}\ee
The general solution of (\ref{f}) is: \bea f(h)&=&\lambda
h^{\f{\g{3_{0}}}{\b{0}}}\Big[1+\Big(\f{\g{3_{1}}}{\b{0}}-\f{\g{3_{0}}\b{1}}{\b{0}^{2}}\Big)h
+\Big(\f{\g{3_{2}}}{2\b{0}}-\f{\g{3_{1}}\b{1}}{2\b{0}^{2}}+\f{\g{3_{0}}}{2\b{0}}(\f{\b{1}}{\b{0}})^{2}-\f{\g{3_{0}}\b{2}}{2\b{0}^2}\nonumber\\
&&+\f{\g{3_{1}}^{2}}{2\b{0}^{2}}-\f{\g{3_{1}}\g{3_{0}}\b{1}}{\b{0}^{3}}+\f{\g{3_{0}}^{2}\b{1}^{2}}{2\b{0}^{4}}\Big)h^2+\ldots\Big],\label{solf}
\eea with $\lambda$ a constant that determines the overall wave
function renormalization. One can easily check the scale and
scheme independence of $\widehat{D}$. From \cite{cr00} and
(\ref{solf}) we find for $\widehat{Z}^{-1}(-q^2)$ (cf.
(\ref{gluonprop})), with $\mu^2=q^2$  and in the $\l{MS}$
scheme:\be
\widehat{Z}^{-1}(-q^2)=\lambda^{-1}h^{-\f{13}{22}}(1+h[-\f{25085}{2904}]
+h^2[-\f{412485993}{1874048}+\f{9747}{352}\zeta_{3}]+\ldots)\label{Zpert}.\ee
(Unfortunately we can only determine ${\widehat{Z}}^{-1}$ up to
second order since for the third order result one needs, besides
the known third order coefficient for $Z^{-1}$ and the four loop
$\beta$-coefficient also the four loop $\g{3}$-coefficient, which
is not available at the moment. As a consequence we are not able
to perform an error estimation.) The 2 loop $y$-expansion result
for ${\widehat{Z}}^{-1}$ is now obtained from (\ref{expansion}),
(\ref{facc}) and (\ref{Zpert}). The 2 loop MOM scheme result is
found as $\lambda^{-1}{h^{(2)}}^{-\f{13}{22}}$ where $h^{(2)}$ is
the solution of (\ref{coupling}) with \be
\b{2}^{\textrm{MOMz}}=\f{105708585}{29744}-\f{107217}{208}\zeta_{3}\qquad
\textrm{and}
\qquad\Lambda_{\textrm{MOMz}}=\Lambda_{\l{MS}}\exp^\f{25085}{37752}.\ee
\begin{figure}
\begin{center}
\includegraphics[angle=-90,width=10cm]{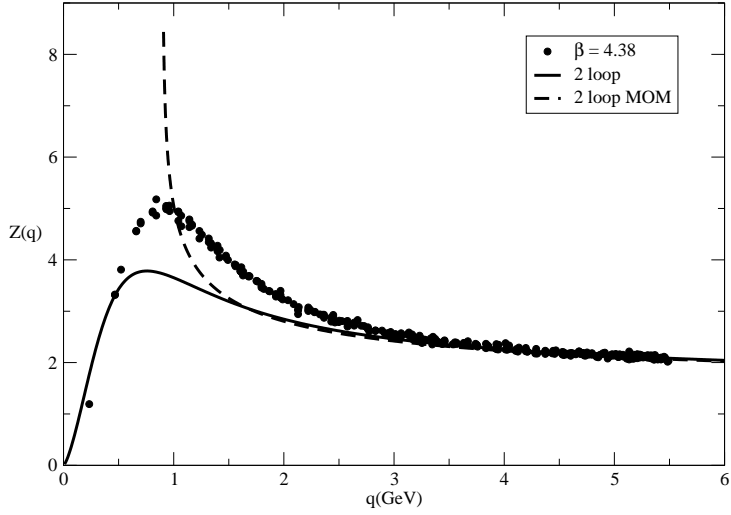}
\caption{Lattice result \cite{bbld01} for the gluon propagator
($q^2\times D(q^2)$) with the two loop results from the
$y$-expansion and the MOM scheme, $\Lambda_{\l{MS}}$=237Mev.}
\label{gluon}
\end{center}
\end{figure}
The 2-loop results for $Z(q^2)$ (Euclidean momentum) are shown
together with a lattice result from \cite{bbld01} in fig.
\ref{gluon}. We now had two fit two things: the scale
$\Lambda_{\l{MS}}$ and the relative wave function renormalization
$\lambda$. Again, we choose the triple gluon vertex value (237
Mev) for $\Lambda_{\l{MS}}$. $\lambda$ is simply determined by
fitting the tail of the 2-loop results on the tail of the lattice
result (at about 5.5 Gev). The overall agreement of our result
with the lattice is similar as for the vertices. In the deep
IR-region, however, there is a discrepancy: in \cite{bbld01} it is
argued, by extrapolation to infinite lattice-volume, that the zero
momentum gluon propagator is finite while we find a singular zero
momentum propagator. Indeed, from the IR-behavior of $y$
(\ref{ylow}) and the expansion for ${\widehat{Z}}^{-1}$
(\ref{Zpert}) one easily obtains the IR-behavior of $D(q)$: \be
D(q)\stackrel{q\rightarrow
0}{\sim}\f{y(q)^{\f{\g{3_{0}}}{\b{0}}}}{q^2}\stackrel{q\rightarrow
0}{\sim}q^{2\f{\b{0}\g{3_{0}}-\b{1}}{\b{1}}}=q^{-\f{61}{102}}. \ee
So our zero momentum result is still singular, although the
singularity is much weaker then the tree level (1/$q^2$) one. We
stress that this specific power behavior will not be altered by
higher loop corrections.
\subsection{The ghost propagator}
The calculation of the ghost propagator is completely similar as
for the gluon propagator. Again we define the scale and scheme
invariant propagator \be
\widehat{G}^{ab}(q)\equiv-\delta^{ab}f_{g}(h)G(q^2)
\equiv-\delta^{ab}\f{\widehat{Z}_{g}(q^2)}{q^2}. \ee From
\cite{cr00} one now arrives at:\be
\widehat{Z}_{g}^{-1}(-q^2)=\lambda_{g}^{-1}h^{-\f{9}{44}}(1+h[-\f{5271}{1936}]
+h^2[-\f{615512003}{7496192}+\f{5697}{704}\zeta_{3}]+\ldots)\label{Zgpert}.\ee
For the three loop MOM $\beta$-coefficient and the
$\Lambda$-parameter we get:\be \b{2}^{\textrm{MOMgh}}=
\f{653203}{176}-\f{6963}{16}\zeta_{3}\qquad \textrm{and}
\qquad\Lambda_{\textrm{MOMgh}}=\Lambda_{\l{MS}}\exp^\f{1757}{2904}.\ee
The two loop results for the Euclidean propagator are plotted
together with the lattice results from \cite{ss96} in fig.
\ref{ghost}. Again we have set $\Lambda_{\l{MS}}$=237 Mev and
$\lambda_{g}$ was determined by fitting the two loop results on
the lattice results at the highest lattice momentum ($\approx 5.5$
Gev, not shown in the fig.). Notice the Landau-pole for the MOM
result. The agreement of our result with the lattice results is
satisfying, apart from the strange single data point at the lowest
lattice momentum. For the IR-behavior of our result we now find:
\be G(q)\stackrel{q\rightarrow 0}{\sim}q^{-\f{103}{68}}~~,\ee
which is more singular then the gluon propagator but less singular
then the tree level result. Although this IR-behavior is
consistent with \cite{ss96}, we should remark that other lattice
studies \cite{nfy01,cmz01} predict a more singular behavior.
\begin{figure}
\begin{center}
\includegraphics[angle=-90,width=10cm]{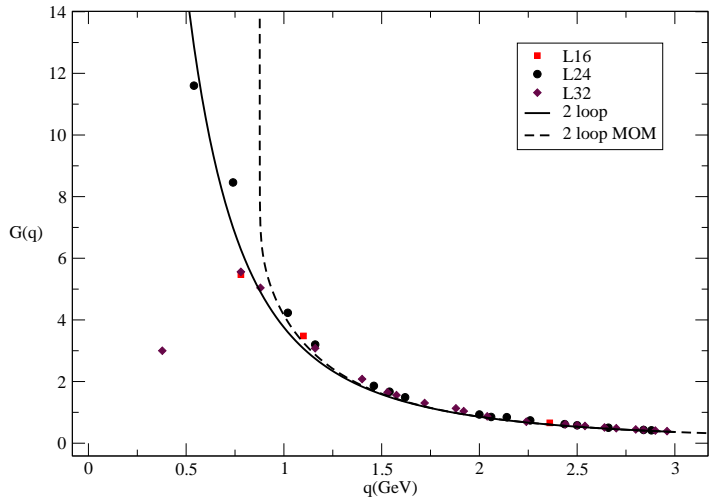}
\caption{Lattice result taken from fig.1 in \cite{ss96} (with
$a^{-1}$=2 Gev) for the ghost propagator with the two loop results
from the $y$-expansion and the MOM scheme,
$\Lambda_{\l{MS}}$=237Mev.} \label{ghost}
\end{center}
\end{figure}

\section{Conclusion}
We have presented an alternative perturbative expansion for QCD
with only one redundant parameter, fixed by a facc, and the
unexpected feature of IR-finite results. The reason behind this
was found to be a universal power behavior of the running
expansion parameter $y$. For zero flavors there is a qualitative
agreement with the lattice data, comparable with the
Schwinger-Dyson results \cite{as01}.

We do not expect our result to cover all the physics in the
intermediate energy region and in the IR region. It is immediately
clear for example, that the $y$-expansion for non-zero quark
flavor will still respect the chiral symmetry, so additional
non-perturbative corrections are definitely needed. However, the
fact that our expansion gives finite results in the whole range of
energies, seems to make it a better (then ordinary perturbation)
framework to start from, if one wants to estimate the true
non-perturbative corrections. In the future we will calculate such
typical sum-rule quantities as current-current 2-point functions.
Other possible applications are the calculation of experimental
cross-sections, starting from the ordinary perturbation theory
results.

\subsection*{Acknowledgements}We would like to thank the lattice
community for providing us with some of their data. In particular
we would like to thank P.Bowman, O.Pene, C.Pittori, J.Micheli and
J.Skullerud. We also thank K.Chetyrkin for giving his
(perturbative) data in several output formats. Finally we thank
O.Bernus for the helping hand on the graphics.


\begin{thebibliography}{10}

\bibitem{s81}
P.~M. Stevenson.
\newblock {\em Phys. Rev.}, D23:2916, 1981.

\bibitem{g84}
G.~Grunberg.
\newblock {\em Phys. Rev.}, D29:2315, 1984.

\bibitem{ss97}
D.~V. Shirkov and I.~L. Solovtsov.
\newblock {\em Phys. Rev. Lett.}, 79:1209--1212, 1997.

\bibitem{b98}
M.~Beneke.
\newblock {\em Phys. Rept.}, 317:1--142, 1999.

\bibitem{ms94}
A.~C. Mattingly and P.~M. Stevenson.
\newblock {\em Phys. Rev.}, D49:437--450, 1994.

\bibitem{svz79}
Mikhail~A. Shifman, A.~I. Vainshtein, and Valentin~I. Zakharov.
\newblock {\em Nucl. Phys.}, B147:385--447, 1979.

\bibitem{kss84}
J.~Kubo, S.~Sakakibara, and P.~M. Stevenson.
\newblock {\em Phys. Rev.}, D29:1682, 1984.

\bibitem{cr00}
K.~G. Chetyrkin and A.~Retey.
\newblock {\em arXiv}, hep-ph/0007088, 2000.

\bibitem{bbr00}
P.~Boucaud et~al.
\newblock {\em JHEP}, 04:006, 2000.

\bibitem{blm98}
P.~Boucaud, J.~P. Leroy, J.~Micheli, O.~Pene, and C.~Roiesnel.
\newblock {\em JHEP}, 10:017, 1998.

\bibitem{vvl97}
T.~van Ritbergen, J.~A.~M. Vermaseren, and S.~A. Larin.
\newblock {\em Phys. Lett.}, B400:379--384, 1997.

\bibitem{cg79}
William Celmaster and Richard~J. Gonsalves.
\newblock {\em Phys. Rev.}, D20:1420, 1979.

\bibitem{s98}
Jonivar Skullerud.
\newblock {\em Nucl. Phys. Proc. Suppl.}, 63:242--244, 1998.

\bibitem{skw01}
Jonivar Skullerud, Ayse Kizilersu, and Anthony~G. Williams.
\newblock {\em Nucl. Phys. Proc. Suppl.}, 106:841--843, 2002.

\bibitem{bbld01}
Frederic D.~R. Bonnet, Patrick~O. Bowman, Derek~B. Leinweber,
Anthony~G.
  Williams, and James~M. Zanotti.
\newblock {\em Phys. Rev.}, D64:034501, 2001.

\bibitem{ss96}
H.~Suman and K.~Schilling.
\newblock {\em Phys. Lett.}, B373:314--318, 1996.

\bibitem{nfy01}
Hideo Nakajima, Sadataka Furui, and Azusa Yamaguchi.
\newblock {\em Nucl. Phys. Proc. Suppl.}, 94:558--561, 2001.

\bibitem{cmz01}
Attilio Cucchieri, Tereza Mendes, and Daniel Zwanziger.
\newblock {\em Nucl. Phys. Proc. Suppl.}, 106:697--699, 2002.

\bibitem{as01}
Reinhard Alkofer and Lorenz von Smekal.
\newblock {\em Phys. Rept.}, 353:281, 2001.

\end{thebibliography}

\end{document}